\documentclass[authorversion, sigconf]{acmart}
\AtBeginDocument{%
  \providecommand\BibTeX{{%
    \normalfont B\kern-0.5em{\scshape i\kern-0.25em b}\kern-0.8em\TeX}}}

\copyrightyear{2024}
\acmYear{2024}
\setcopyright{rightsretained}
\acmConference[AVI 2024]{International Conference on Advanced Visual Interfaces 2024}{June 3--7, 2024}{Arenzano, Genoa, Italy}
\acmBooktitle{International Conference on Advanced Visual Interfaces 2024 (AVI 2024), June 3--7, 2024, Arenzano, Genoa, Italy}\acmDOI{10.1145/3656650.3656690}
\acmISBN{979-8-4007-1764-2/24/06}

\newcommand{\commentout}[1]{}

\usepackage{color}
\usepackage{setspace}
\definecolor{Orange}{rgb}{1,0.5,0}
\definecolor{DarkGreen}{rgb}{0,0.5,0}
\definecolor{Purple}{rgb}{0.7,0,0.7}
\definecolor{Blue}{rgb}{0.2,0.2,0.8}
\definecolor{Red}{rgb}{1.0,0.0,0.0}
\definecolor{Brown}{rgb}{0.7,0.4,0.1}

\usepackage{multicol}
\usepackage{booktabs}
\usepackage{enumitem}

\usepackage{array}
\usepackage{longtable}
\usepackage{tabularray}

\begin{document}

%%
%% The "title" command has an optional parameter,
%% allowing the author to define a "short title" to be used in page headers.
\title{Pinching Tactile Display: A Cloth that Changes Tactile Sensation \\by Electrostatic Adsorption}

%%
%% The "author" command and its associated commands are used to define
%% the authors and their affiliations.
%% Of note is the shared affiliation of the first two authors, and the
%% "authornote" and "authornotemark" commands
%% used to denote shared contribution to the research.
\author{Takekazu Kitagishi}
\orcid{1234-5678-9012}
\affiliation{%
  \institution{The University of Tokyo}
  \institution{ZOZO Research}
  \city{Tokyo}
  \country{Japan}
}
% \email{kitagishi-takekazu588@g.ecc.u-tokyo.ac.jp}
\email{ktgs@acm.org}

\author{Hirotaka Hiraki}
\orcid{1234-5678-9012}
\affiliation{%
  \institution{The University of Tokyo}
  \city{Tokyo}
  \country{Japan}
}
\email{hiraki-uts1@g.ecc.u-tokyo.ac.jp}

\author{Hiromi Nakamura}
\affiliation{%
  \institution{The University of Tokyo}
  \city{Tokyo}
  \country{Japan}
}
\email{
% hiromi.nakamura@iii.u-tokyo.ac.jp
% hirominakamura.b@gmail.com
hirominakamura@sigchi.org
}

\author{Yoshio Ishiguro}
\affiliation{%
  \institution{The University of Tokyo}
  \city{Tokyo}
  \country{Japan}
}
\email{ishiy@acm.org}

\author{Jun Rekimoto}
\affiliation{%
  \institution{The University of Tokyo}
  \institution{Sony CSL Kyoto}
  \city{Tokyo / Kyoto}
  \country{Japan}
}
\email{rekimoto@acm.org}
%%
%% By default, the full list of authors will be used in the page
%% headers. Often, this list is too long, and will overlap
%% other information printed in the page headers. This command allows
%% the author to define a more concise list
%% of authors' names for this purpose.
\renewcommand{\shortauthors}{Takekazu Kitagishi, Hirotaka Hiraki, Hiromi Nakamura, Yoshio Ishiguro and Jun Rekimoto}

%%
%% The abstract is a short summary of the work to be presented in the
%% article.
\begin{abstract}
    Haptic displays play an important role in enhancing the sense of presence in VR and telepresence.
Displaying the tactile properties of fabrics has potential in the fashion industry, but there are difficulties in dynamically displaying different types of tactile sensations while maintaining their flexible properties.
The vibrotactile stimulation of fabrics is an important element in the tactile properties of fabrics, as it greatly affects the way a garment feels when rubbed against the skin.
To dynamically change the vibrotactile stimuli, many studies have used mechanical actuators.
However, when combined with fabric, the soft properties of the fabric are compromised by the stiffness of the actuator.
In addition, because the vibration generated by such actuators is applied to a single point, it is not possible to provide a uniform tactile sensation over the entire surface of the fabric, resulting in an uneven tactile sensation.
In this study, we propose a Pinching Tactile Display: a conductive cloth that changes the tactile sensation by controlling electrostatic adsorption.
By controlling the voltage and frequency applied to the conductive cloth, different tactile sensations can be dynamically generated.
This makes it possible to create a tactile device in which tactile sensations are applied to the entire fabric while maintaining the thin and soft characteristics of the fabric.
As a result, users could experiment with tactile sensations by picking up and rubbing the fabric in the same way they normally touch it. 
This mechanism has the potential for dynamic tactile transformation of soft materials.

\end{abstract}


%%
%% Keywords. The author(s) should pick words that accurately describe
%% the work being presented. Separate the keywords with commas.

%% A "teaser" image appears between the author and affiliation
%% information and the body of the document, and typically spans the
%% page.
\begin{teaserfigure}
  \includegraphics[width=\textwidth]{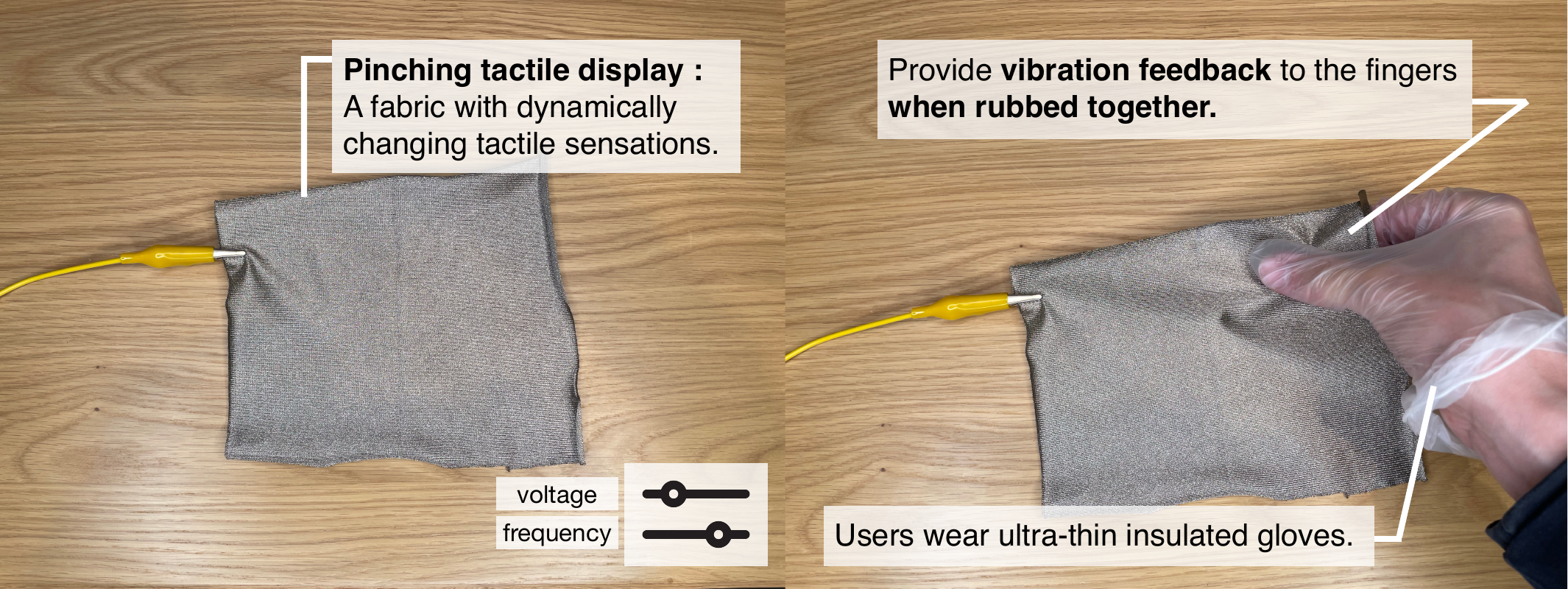}
  \caption{Overview of the Pinch Tactile Display. The fabric changes its tactile sensation by manipulating the electrostatic force between the conductive fabric and the user's finger. Users interact with it by wearing gloves and rubbing them together. The texture variation is achieved by adjusting the voltage and frequency, with the voltage modulated by the power supply and the frequency controlled by Arduino. The inherent thinness and softness of the fabric allows for a natural touch experience, enhancing immersive interactions in virtual reality environments.}
  % \Description{teaser. explain how to use it. explain how it works. explain what is new.}
  \label{fig:teaser}
\end{teaserfigure}

% \received{20 February 2007}
% \received[revised]{12 March 2009}
% \received[accepted]{5 June 2009}

%%
%% This command processes the author and affiliation and title
%% information and builds the first part of the formatted document.
\maketitle

% \footnotetext[1]{\copyright{} 2024 ACM. This is the author's version of the work. It is posted here for your personal use. Not for redistribution. The definitive Version of Record was published in SourcePublication, http://dx.doi.org/10.1145/number.}

% "© Owner/Author(s) | ACM, 2024. This is the author's version of the work. It is posted here for your personal use. Not for redistribution. The definitive Version of Record was published in AVI 2024, http://dx.doi.org/10.1145/3656650.3656690."

% \footnotetext[1]{
% \copyright{} Owner/Author(s) | ACM, 2024. This is the author's version of the work. It is posted here for your personal use. Not for redistribution. The definitive Version of Record was published in AVI 2024, \url{http://dx.doi.org/10.1145/3656650.3656690}
% }
% \titlenote{%
%   \copyright{} Owner/Author(s) | ACM, 2024. This is the author's version of the work. 
%   It is posted here for your personal use. Not for redistribution. 
%   The definitive Version of Record was published in AVI 2024, 
%   \url{http://dx.doi.org/10.1145/3656650.3656690}.}

% \input{1.introduction}
% \input{2.relatedwork}
\section{introduction}

Haptic displays play an important role in enhancing the sense of presence in VR and telepresence. They enhance the realism of objects by presenting the texture of objects that are distant or out of place. As a result, it can improve the sense of presence and immersion when manipulating virtual reality spaces and create a sense of trust in remote objects.

In order to improve the sense of presence in VR and telepresence scenarios, attempts have been made to represent the tactile sensations of various objects around us in our daily lives.
In the process of trying to reproduce the tactile sensations of such everyday objects, reproducing the tactile sensations of soft objects is challenging because the shape of soft objects can be changed more easily by pushing or pulling than that of hard objects.

Therefore, the aim of this study is to present the tactile sensation of soft cloth.
Cloth is used daily for clothing, bedding, sofas, chairs, etc.
The tactile sensation of these materials is something we experience on a daily basis, and if it can be changed according to an individual's preferences, it is expected to make the individual's daily life more comfortable.
Therefore, the presentation of soft fabric tactile sensations can be expected to have an impact on the fashion and furniture industries.
For example, if online shoppers can check the feel of a product before purchasing it, it will reduce the experience of the purchased product not feeling like what they wanted to wear, thereby improving the quality of online shopping.

The vibrotactile stimuli that people feel when the fabric rubs against their skin are important to the texture of the fabric. When creating a device that dynamically changes and presents vibrotactile sensations, it is common to use vibration generated by a mechanical actuator such as a motor~\cite{choi2012vibrotactile, wiertlewski2011reproduction}.
However, when a mechanical actuator is applied to cloth and a mechanical actuator is used to generate the vibration stimulus produced when the cloth is picked up and rubbed with the fingers, the softness of the cloth is reduced because the actuator itself is stiff.
In addition, the vibration generated by the mechanical actuator attenuates as it moves away from the area being vibrated, making it impossible to provide a uniform tactile sensation over the entire surface of the cloth. This reduces the uniform weave of the cloth and the property that the entire cloth has a uniform tactile feel to some extent.

In this paper, we propose a Pinching Tactile Display, a cloth that changes tactile sensation by electrostatic adsorption.
It has a mechanism to transform the tactile sensation by controlling the electrostatic force between the conductive fabric and the finger without using a mechanical actuator. It inherits the thin and soft property of the cloth. It has uniform tactile sensations over the whole cloth, which allows us to obtain tactile sensations by pinching and rubbing the cloth as we normally do when testing the feel of the cloth. This increases the reality of how we interact with the fabric.

Users wear ultra-thin insulated gloves and touch the conductive fabric to prevent electric shock and to feel the delicate tactile sensations of the cloth. Users can experience multiple tactile sensations of cloth with a single device.
We confirmed that this mechanism changes the tactile sensation of the fabric. We also confirmed that different tactile sensations can be generated by switching the voltage and frequency.

Our contributions include the following: 
\begin{itemize}
  \item We have developed a technique that allows us to change the tactile feel of the cloth while maintaining its thin and soft nature.
  \item It allows users to pinch and rub the conductive fabric to check its tactile feel, just as they would normally check the tactile feel of a fabric.
  \item By controlling the voltage and frequency of the electricity applied to the conductive cloth, we can dynamically change the tactile feel of the cloth.
\end{itemize}

\section{related work}

\subsection{Haptic Display for Hard objects}
The most common method of tactile transmission is the presentation of vibration by mechanical vibration.~\cite{choi2012vibrotactile, wiertlewski2011reproduction} introduced eccentric rotating mass motors,~\cite{spirkovska2005summary, wiertlewski2011reproduction} introduced voice coils,~\cite{cascio2001temporal} introduced piezo electric actuators to represent the texture of fabrics. However, when combined with fabric, the soft properties of the fabric are compromised by the stiffness of the actuator. In addition, because the vibration generated by such actuators is applied to a single point, it is not possible to provide a uniform tactile sensation over the entire surface of the fabric, resulting in an uneven tactile sensation. 

There is a study that uses a touch panel to convey the tactile sensation of the fabric while complementing it with visuals.
TPaD~\cite{4145211} and ShiverPaD~\cite{5438990} is based on the theory, design and construction of tactile displays that create texture sensations through changes in surface friction, known as the squeeze film effect. Using ultrasonic frequency and low amplitude vibrations between two flat plates, a squeezed film of air is created between the surfaces, reducing friction. 
TeslaTouch~\cite{bau2010teslatouch} is a technique for creating virtual textures by controlling the shear force on a touch panel. It involves a simple structure where high voltage is applied and an insulator is placed between the user and the panel, creating a shear force when touched. This allows the creation of thin devices that can change their feel when rubbed together. However, despite its simplicity, TeslaTouch targets touch panels that are inherently hard, creating challenges in accurately representing the feel of the material when rubbed together.
The touch panel is hard, which limits its ability to convey the soft tactile sensation of the fabric.

\begin{figure}[t]
  \centering
  \includegraphics[width=\linewidth]{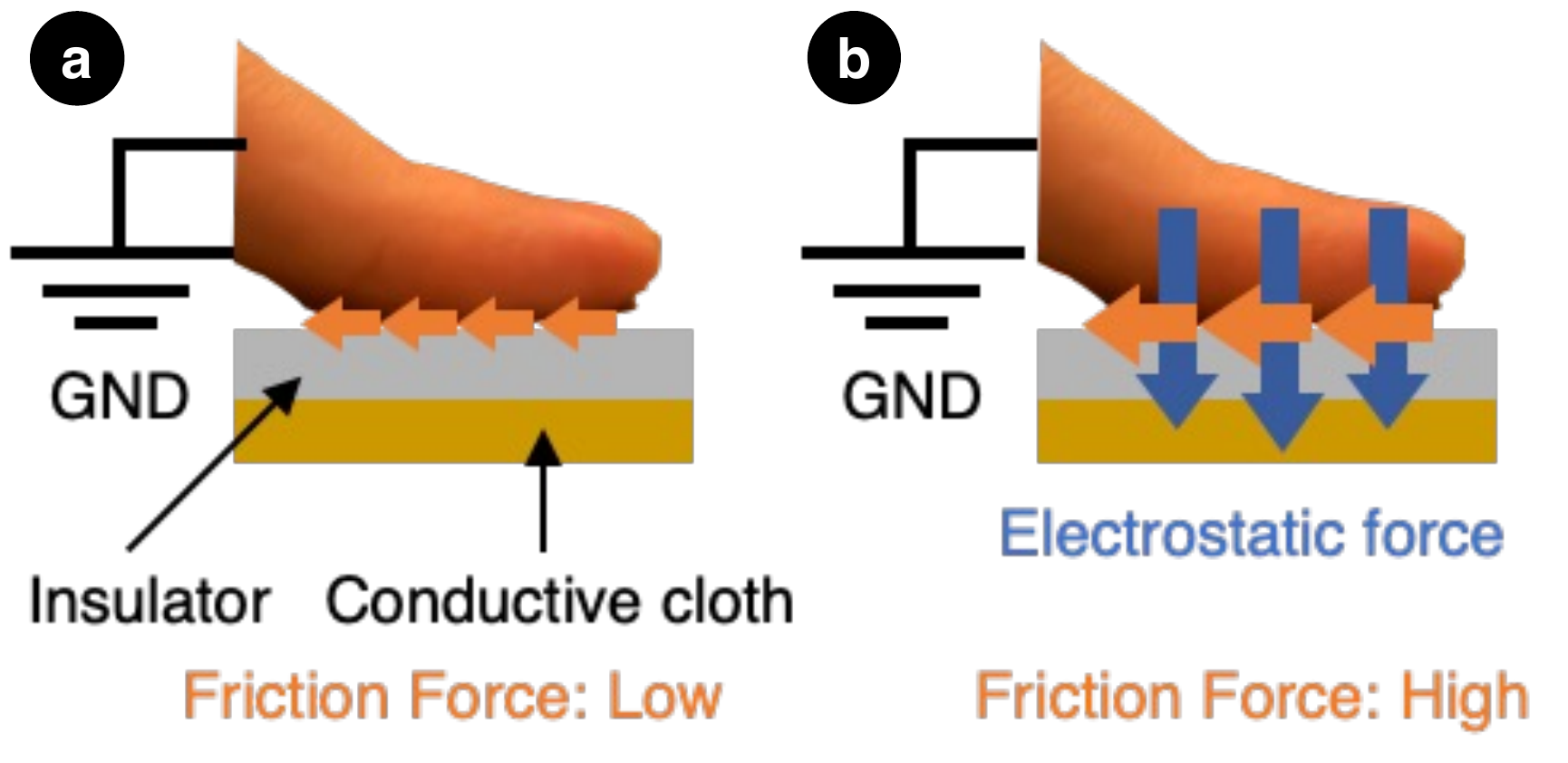}
  % \caption{Principle of the Pinching Tactile Display. (a) Without Voltage. (b) With Voltage.}
  \caption{Principle of the Pinching Tactile Display. (a) Without Voltage (b) With Voltage. Applying a voltage to a conductive fabric creates a charge distribution that generates an electrostatic force on the skin and a frictional force. By periodically varying the frictional force, the skin in contact is periodically deformed. As a result, the user feels vibratory stimulation.}
  % \Description{A woman and a girl in white dresses sit in an open car.}
  \label{fig:mechanism_of_electrovibration}
\end{figure}

\subsection{Haptic Display for Soft Objects}
% Some devices are designed to provide soft tactile sensations by controlling the shape of soft objects.
% ~\cite{10.1145/3586183.3606765} provides passive tactile sensations by exploiting the attraction between ferromagnetic yarns and permanent magnets, both of which are seamlessly integrated into knitted fabrics. In~\cite{10.1145/3472749.3474798}, the surface structure of the fabric is 3D scanned and the same structure is 3D printed to provide tactile sensations.~\cite{10.1145/3290605.3300479} present a soft texture using a hair structure.
% There is only one type of tactile sensation that can be presented by a single device, and it takes time to create that device, and the tactile sensation cannot be changed dynamically.
% Ultrasonic~\cite{Watanabe1995-sf, Wiertlewski2015-cb} displays tactile sensation of soft objects and it can dynamically change it.
% However, they often require expensive or large setups, which does not fit our goal of providing a simple experience of the tactile sensation of soft objects.
% Based on the above, this study proposes a mechanism to dynamically change the tactile sensation of soft fabrics without expensive or large setups.
One way to present soft tactile sensations is to change the shape of soft objects.
~\cite{10.1145/3586183.3606765} uses the attraction between ferromagnetic yarns and permanent magnets seamlessly integrated into knitted fabrics to provide a passive tactile sensation. In addition, ~\cite{10.1145/3472749.3474798} provides tactile sensations by 3D scanning the surface structure of fabrics and 3D printing the same structures. They envision controlling the shape of a soft object and combining it with vision to provide a soft tactile sensation, but only one type of tactile sensation can be presented with a single device, the device takes time to create, and the tactile sensation cannot be dynamically changed.

Some research displays tactile sensations of soft objects by allowing the user to touch a real piece of fabric. HapticRevolver~\cite{10.1145/3173574.3173660} is an actuator wheel that rises under the finger and makes contact with a virtual surface. Haptic Palette~\cite{10.1145/3393914.3395870} extends this wheel-based actuator with visual enhancements on top of the physical texture in virtual reality environments, allowing the user to experience mixed material perception. Telextiles~\cite{10.1145/3586183.3606764} uses a roller-type device to present the closest tactile match to a remote fabric.
These devices are limited in the number of fabrics that can be attached to them, and the tactile sensation of the fabrics does not change continuously and therefore cannot be dynamically changed.

Ultrasonic~\cite{Watanabe1995-sf, Wiertlewski2015-cb} can display and dynamically change the tactile sensation of soft objects.
However, they often require expensive or large equipment and are not suitable for the purpose of easily experiencing the tactile sensation of soft objects.

In view of the above, this study proposes a mechanism to dynamically change the tactile sensation of soft cloth without using expensive or large-scale devices.

\subsection{Electrovibration}
Electrovibration is a phenomenon that allows us to create tactile sensations by controlling the {\it electrostatic friction} between a dielectric material and the user's fingers. The discovery of electrovibration dates back to 1954, when it was discovered by accident. Mallinckrodt et al. reported that dragging a dry finger across a conductive surface covered with a thin insulating layer and stimulated with a 110 V signal resulted in a characteristic ``rubbery'' sensation~\cite{doi:10.1126/science.118.3062.277}. This sensation was explained by the theory that the insulating layer of the dry outer skin acts as the dielectric layer of a capacitor. In this model, the conductive surfaces and fluids within the finger tissue represent the two opposing plates of the capacitor.
While this force is too weak to be felt when the finger is static, it significantly modulates the friction between the surface and the skin of the moving hand, creating the unique rubbery sensation associated with electrovibration.

\begin{figure}[t]
  \centering
  \includegraphics[width=\linewidth]{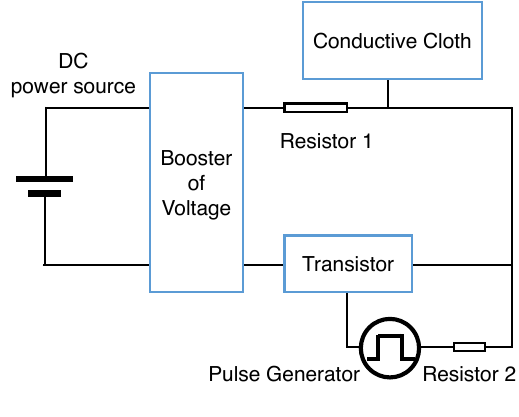}
  % \caption{System Overview}
  \caption{System Overview. When the switch is turned on, current flows, causing a voltage drop across Resistor 1, resulting in 0V across the Conductive Cloth. Conversely, when the switch is off, no current flows, resulting in no voltage drop across Resistor 1 and a high voltage across the Conductive Cloth. The switching mechanism is controlled by the pulse generator and a transistor. The switch is activated (turned on) when voltage is applied to the transistor and deactivated (turned off) when no voltage is applied. This configuration allows a periodic voltage to be applied to the conductive cloth.}
  % \Description{A woman and a girl in white dresses sit in an open car.}
\end{figure}

\section{System Configuration}
We propose Pinching Tactile Display, a cloth that changes tactile sensation by electrostatic adsorption. By changing the voltage and frequency of the square wave applied to the conductive cloth, we can control the electrostatic adhesion force between the cloth and the finger to generate multiple tactile sensations of the cloth. The user puts on the gloves and rubs his or her fingers against the cloth. They can experience multiple tactile sensations with a single device. The gloves are ultra-thin, non-conductive and act as an insulator, preventing electricity from flowing from the fabric to the fingers while minimizing the space between the fingers and the fabric, allowing the user to feel the subtle tactile sensations of the fabric. We can adjust the voltage applied to the conductive fabric by adjusting the power supply voltage, and the frequency by controlling the switching operation of the transistors. This system provides tactile sensation throughout the conductive cloth while maintaining the thin and soft characteristics of the cloth. As a result, users can explore the tactile sensation by pinching the cloth and rubbing it, as they normally do when touching cloth.

% \subsection{Experimental Design}

% \subsection{Mechanism of change of tactile sensation}

\subsection{Principle of Pinching Tactile Display}

% Figure~\ref{fig:mechanism_of_electrovibration} illustrates the principle of the Pinch Tactile Display. This system uses the phenomenon of electrovibration. In the absence of high voltage, no external force is exerted on a finger stroking the tactile display. As a result, users perceive a soft sensation from the insulator and cloth, as shown in Figure~\ref{fig:mechanism_of_electrovibration}. When a voltage is applied to the conductive cloth, the finger connected to GND becomes negatively charged and the electrodes connected to the high voltage become positively charged. This charge induces an electrostatic force on the skin. The periodic variation of the frictional force causes periodic deformation of the contacting skin. As a result, the user experiences a vibrational stimulus.
Figure~\ref{fig:mechanism_of_electrovibration} illustrates the principle of the Pinch Tactile Display, which uses the phenomenon of electrovibration. This display consists of a conductive cloth and an insulator, such as a PVC glove. We connected the conductive fabric to a high voltage supply. Without high voltage, no external force is applied to a finger moving over the tactile display. Consequently, the user perceives the original surface of the conductive cloth, as shown in Figure~\ref{fig:mechanism_of_electrovibration}(a). When voltage is applied to the conductive cloth, the finger becomes negatively charged, while the conductive cloth connected to the high voltage becomes positively charged. This charge distribution creates an electrostatic force on the skin, resulting in a frictional force, as shown in Figure~\ref{fig:mechanism_of_electrovibration}(b). By periodically changing the frictional force, the contacting skin deforms periodically. As a result, the user perceives a vibratory stimulus.

The attractive force and the resulting frictional force can be succinctly expressed as follows:
\begin{equation}
    F = A \frac{\epsilon\epsilon_0}{2} \left(\frac{V(t)}{d}\right)^2 
    \label{eq:electrostatic force}
\end{equation}
\begin{equation}
    F' = \mu F = \mu A \frac{\epsilon\epsilon_0}{2} \left(\frac{V(t)}{d}\right)^2
    \label{eq:electrostatic_friction}
\end{equation}
where $F$ is the attractive force, $F'$ is the frictional force, $\epsilon_{0}$ is the vacuum permeability, $\epsilon$ is the relative permeability of the insulator, $A$ is the contact area, $V(t)$ is the applied voltage, $d$ is the thickness of the insulator, and $\mu$ is the coefficient of friction. This equation implies that the voltage, the time period (frequency), and the distance between the finger and the conductive cloth significantly affect the electrostatic friction.

\subsection{Controlling Tactile Sensation of Cloth}
% \section{Tactile Modification}

This system alters tactile sensations by varying the magnitude and frequency of the square-wave voltage applied to the conductive cloth.
The proposed mechanism varies the voltage applied to the Conductive Cloth by changing the voltage of the DC power source input to the Voltage Booster. The voltage on the Conductive Cloth changes depending on the voltage output by the DC power source and the function by which the Booster of Voltage transforms the input voltage into an output voltage.

The proposed mechanism uses the switching action of a transistor to adjust the on/off state of the voltage connected to the conductive cloth, thereby applying a voltage in the form of square waves. The switching of the transistor is controlled by a pulse generator. When the pulse generator applies voltage, the switch turns on; when no voltage is applied from the pulse generator, the switch turns off. When the switch is on, a high voltage increased by the voltage booster is applied to the conductive cloth; when the switch is off, no voltage is applied to the cloth. By equalizing the on and off times, square waves are generated. The frequency of the voltage applied to the Conductive Cloth changes according to the duration of the on and off states of the switch.

% This system changes Voltage and frequency of the 

% The amplitude and frequency of the square wave voltage applied to the conductive fabric changes the tactile sensation of the conductive fabric. We can change the voltage by changing the voltage of the circuit's stabilized power supply. The DC-DC converter amplifies the electricity generated by the stabilized power supply, which increases the voltage to approximately 100V - 300V for the generation of tactile sensations. The characteristics of the DC-DC converter determine how the input voltage is converted to the desired output voltage. We used Takasago Seisakusho's LX-2-035-1B compact switching power supply for the stabilized power supply and XP Power's DC-DC converter for the voltage boost.

% The system controls the frequency of the square wave voltage by adjusting the frequency settings on an Arduino. The mechanism behind this is the use of a MOS-FET for switching within the circuit. When the Arduino applies voltage, the MOS-FET switch turns on, and conversely, when no voltage is applied, the switch turns off. By alternating between 5V and 0V with a constant period, we ensure that the voltage takes the form of a square wave. This square wave is essential for creating the electrovibration effect, as it allows precise control of the electrostatic force generated between the conductive cloth and the user's finger.

\subsection{Conductive Fabric and Thin Glove for Insulator}
The user wears gloves to minimize the distance between the fingers and the fabric while preventing electricity from flowing from the fabric to the fingers. As shown in the equation ~\ref{eq:electrostatic_friction}, distance is an important factor that is the square of the equation, and the thinness of the material is an important factor in sensing tactile changes. Thin gloves also allow the wearer to feel the subtle tactile sensations of the material.
By using the method of putting on the gloves first, the user can finish feeling the insulator before touching the fabric and can concentrate on the feel of the fabric. In addition, the gloves are relatively close to the fingers and do not interfere much with the feel of the fabric. Because any thin, insulating material can be used, they are inexpensive, widely available, and easy to obtain.

To ensure that the voltage spreads evenly over all areas, we used a conductive fabric made of 100\% metal fibers with a uniform surface structure. Like typical fabrics, it has the characteristics of being thin and flexible.

\subsection{Safety Measures}
% To achieve safe operation, the current flowing in the high-voltage part of the tactile sensing device and the circuit connected to it should be set to a very low amount. The current in the high-voltage portion of the circuit is limited to 0.5 mA, which is considered safe for humans \cite{alma991031483099703276}. This amount of current is comparable to the amount of current that flows through a user's hand when using a capacitive touch screen used in a typical smartphone \cite{3m_undated-iu}.

For safety, the current flowing through the high-voltage part of the tactile device and the circuit connected to it was set very low. The current flowing in the high-voltage section was limited to $0.5~mA$ or less, which is considered safe for the human body~\cite{alma991031483099703276}. This amount of current is said to be the same as the amount of current that flows through a user's hand when using a capacitive touch screen used in a typical smartphone~\cite{3m_undated-iu}.

\begin{figure}[t]
  \centering
  \includegraphics[width=\linewidth]{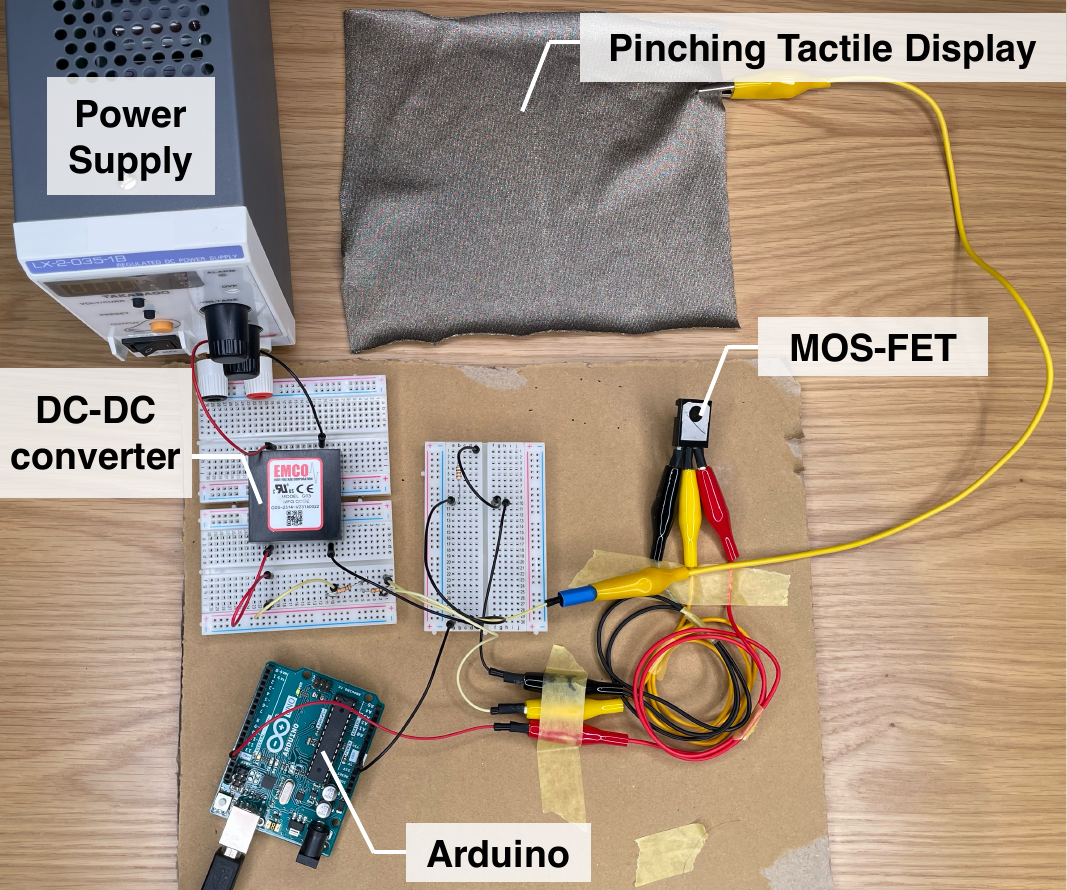}
  \caption{
  % The pinching tactile display configuration, including a conductive cloth, a DC-DC converter, a MOS-FET, Arduino and a power supply.
  System setup for evaluation. The system consists of a conductive cloth, a DC-DC converter for booster of voltage, a MOS-FET for transistor, Arduino for pulse generator and a power supply.
  }
  % \Description{A woman and a girl in white dresses sit in an open car.}
\end{figure}

\section{Evaluation}
To evaluate whether we could realize a cloth that changes its tactile sensation, we conducted a user test. We investigated whether the device's tactile sensation changes by controlling the electrostatic adhesion force and whether the fabric maintains its softness and flexibility even when the tactile sensation changes.

The hypotheses are as follows:

\begin{description}
    \item[H1] The magnitude of the electrostatic force, which pulls the user's finger towards the cloth, varies with the voltage, and the frequency of this pulling force changes with the frequency of the applied signal. Therefore, by switching the voltage and frequency, users can perceive different tactile sensations from the device.

    \item[H2] Since no rigid structures are used for the vibration of the cloth, when a person touches the cloth, the properties of the cloth remain unchanged, and only the tactile sensations related to electrovibration change.
\end{description}

\subsection{Displayed Tactile sensation}
To verify (H1) people can feel different sensations from the device by switching the voltage/frequency, we changed the voltage/frequency applied to the cloth and had subjects touch it. We then asked the subjects whether they display different tactile sensations, how they changed with voltage and frequency respectively and what these sensations were like.

We changed the voltage between $100~V, 200~V$ and $300~V$ and the frequency between $50~Hz, 100~Hz$ and $200~Hz$. Each of these was combined to create a total of $3 \times 3 = 9$ different states. Then we asked subjects in a total of $10$ different states: nine of these and one in which no electricity was applied.

% \subparagraph{\textbf{Change in Tactile Sensation}}
% \paragraph{\textbf{Change in Tactile Sensation}}
% \textbf{Change in Tactile Sensation}
\paragraph{Change in Tactile Sensation}
To check whether this device display different tactile sensations, we asked users how many different tactile sensations they felt after touching the cloth in all states. In order to obtain statistics on how many different tactile sensations were felt out of a total of nine, we calculated the mean and variance of the types of responses given.

% \subparagraph{\textbf{Effects of Voltage and Frequency on Tactile Sensation}}
% \paragraph{\textbf{Effects of Voltage and Frequency on Tactile Sensation}}
% \textbf{Effects of Voltage and Frequency on Tactile Sensation}
\paragraph{Effects of Voltage and Frequency on Tactile Sensation}
To ascertain how their tactile sensations changed with the respective changes in voltage and frequency, we analysed statistically the Likert Scale answers. To determine which voltage, frequency or their alternating factors had a predominant effect on the tactile properties of the cloth, we analysed the responses obtained for a total of $3 \times 3 = 9$ different states, combining three types of voltage ($100~V, 200~V$ and $300~V$) and three types of frequency ($50~Hz, 100~Hz$ and $200~Hz$). Since this obtained data are a two-factor within-participant design with a within-participant factor (voltage) × within-participant factor (frequency), and since normality cannot be assumed for the data obtained, we applied an aligned rank transform (ART) and 
conducted a two-way ANOVA~\cite{10.1145/1978942.1978963}.

% \subparagraph{\textbf{Detailed Property of the Tactile Sensation}}
% \textbf{Detailed Property of the Tactile Sensation}
\paragraph{Detailed Property of the Tactile Sensation}
To confirm the property of the tactile sensations displayed, we asked subjects two questions. First, for all $10$ states of the cloth, we asked subjects to touch the cloth and freely describe their impression of the tactile sensation of the cloth. Second, they touched the cloth and answered a five-point Likert Scale for the four properties: roughness, thickness, stiffness and warmth. Our choice of descriptive term pairs relates to the findings of Soufflet, Calonnier and Dacremont~\cite{soufflet2004comparison} who demonstrate that the most significant scales their participants used to rate the properties of textiles were Stiff-Flexible, Thick-Thin and Soft-Harsh. We calculated the average of the answers for each index for each state of the cloth and visualized to interpret what the respective trends were.

\subsection{Properties as Cloth}
To test the hypothesis (H2) the properties of the fabric remain unchanged even when the electrostatic force is controlled, we varied the voltage and frequency applied to the fabric, as we did in the previous chapter. For each condition, we asked participants whether the tactile sensation of the fabric was acceptable and what type of fabric the tactile sensation resembled.

% \subparagraph{\textbf{Is It Acceptable as a Cloth?}}
% \textbf{Is It Acceptable as a Cloth?}
\paragraph{Is It Acceptable as a Cloth?}
Regarding the acceptability of the tactile sensation as a fabric, we hypothesized that the tactile sensation would be acceptable when the voltage was low, and that the impression of electricity would become stronger and the impression of the fabric would become weaker when the voltage was high. After touching the fabric in each state, we asked participants if the tactile sensation was acceptable as a fabric or if they felt that the fabric had lost its texture. We calculated the mean and variance of the number of acceptable states out of the 10 states of the fabric.

% \subparagraph{\textbf{Tactile Similarity to Familiar Cloth}}
% \textbf{Tactile Similarity to Familiar Cloth}
\paragraph{Tactile Similarity to Familiar Cloth}
Regarding the similarity of the presented tactile sensation to the tactile sensation of a fabric, we hypothesized that it would be similar to the tactile sensation of a fabric with increased friction compared to the original conductive fabric. We selected 16 types of fabrics from a fabric sample book~\cite{Tanaka2009-vp}, carefully choosing fabrics with different materials and folding methods to ensure a variety of textures. As before, we created 10 states of the conductive cloth by applying electricity to it. For each state, we asked the participants to select the fabric that was most similar in tactile sensation from among the 16 types of fabrics. This was done for each participant individually. Next, we tallied and visualized how many times each of the 16 fabrics was selected. We also tallied and visualized how many times each type of tissue was selected for each voltage and frequency category. Through these processes, we considered the characteristics of the fabrics for which similar tactile sensations could be produced with this system.

\subsection{System Setup}

The experimental setup focuses on a system designed to control electrostatic forces on a conductive cloth. The system is designed to generate tactile sensations by manipulating electrostatic forces through variable voltage and frequency inputs.The core component of our system is the conductive cloth, which is essential for applying the desired voltages. We used a conductive cloth measuring approximately $30~cm \times 30~cm$, made of $100~\%$ silver fiber, with an electrical resistance of 1 ohm and an electromagnetic shielding effectiveness of 60DB. The cloth is knitted to provide elasticity, with a weight of $90~g/ m^{2}$ and a thickness of $0.232~mm$.

To control the electrostatic force, the system uses Arduino to generate square waves at $50~Hz, 100~Hz$ and $200~Hz$ in the range of 0V to 5V. A MOS-FET is used to turn off at 0V and turn on at 5V. This setup allows the conductive fabric to be subjected to $100~V, 200~V$ and $300~V$ when the switch is on and 0V when it is off, thereby applying square wave voltages to the conductive fabric at the voltage and frequency. The MOSFET model used is the silicon N-channel MOS (DTMOS-H) TK31N60X. The system also includes Arduino UNO, a DC-DC converter (EMCO G03) and a power supply from Matsusada. The specific resistors and currents used in the system are yet to be determined.

In addition to the conductive fabric, ultra-thin insulating gloves were used in the experiments. These gloves are made of polyvinyl chloride (PVC) with a thickness of $0.035~mm$, measured with a micrometer. The gloves are powder-free on the inside and measure $23.5~cm \times 10.5~cm \times 8.5~cm$. This setup allows a controlled interaction between the user's skin and the electrostatically charged fabric, allowing a detailed study of the tactile sensations produced by different electrostatic conditions.

\begin{figure}[t]
  \centering
  \includegraphics[width=\linewidth]{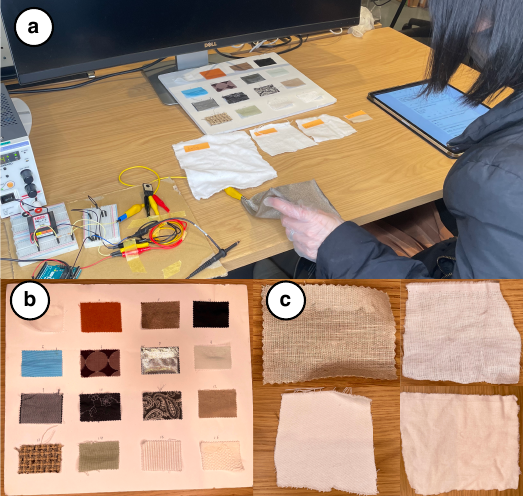}
  % \caption{Experimental Procedure}
  \caption{Experimental procedure. (a) Subjects answered questions about the tactile sensation of the conductive cloth. (b) Subjects selected the most similar tactile sensation cloth among the 16 prepared fabrics. (c) Subjects answered a 5-point Likert scale for roughness, thickness, stiffness, and warmth based on four fabrics: jeans, voile, gauze, and towels.}
  % \Description{A woman and a girl in white dresses sit in an open car.}
\end{figure}

\begin{table}
  \caption{Criteria cloth and their properties.}
  \label{tab:cloth_criteria}
  \begin{tabular}{r r r }
    \toprule
    \textbf{ROUGH} & \textbf{NEUTRAL} & \textbf{SMOOTH} \\
    \midrule
    Jeans   & Voile & Gauze \\
    \midrule
    \textbf{STIFF} & \textbf{NEUTRAL} & \textbf{FLEXIBLE} \\
    \midrule
    Towelling   & Gauze & Voile \\
    \midrule
    \textbf{THICK} & \textbf{NEUTRAL} & \textbf{THIN} \\
    \midrule
    Jeans   & Towelling & Gauze \\
    \midrule
    \textbf{WARM} & \textbf{NEUTRAL} & \textbf{COOL} \\
    \midrule
    Towelling   & Jeans & Voile \\
    \bottomrule
  \end{tabular}
\end{table}

\subsection{Experimental Procedure}
We conducted a user study with six participants, five men and one woman, all in their twenties. We conducted all experiments in accordance with the safety standards approved by the Ethics Research Committee of our institute. We explained the experiment to the participants prior to their participation, and they signed informed consent forms. In addition, we conducted the study in accordance with the ethical standards of the Declaration of Helsinki.

Subjects answered the following question A about the tactile sensation of the cloth without prior application of electricity. The reason for having them touch the cloth first was to confirm the properties of the original conductive cloth by touch and then use it as a basis for the tactile sensation when electricity was applied. Next, three types of voltage ($100~V, 200~V$, and $300~V$) and three types of frequency ($50~Hz, 100~Hz$, and $200~Hz$) were used to create a total of $3 \times 3 = 9$ types of conductive cloth. For each type, the participants answered question A about the tactile sensation of the cloth. To eliminate the influence of the order of touching and the tactile impression of the previously touched cloth, the order in which the cloth was presented in each state was randomized. Finally, participants were asked how many of the nine tactile sensations they thought were manifested in the nine tactile sensations presented.

Question A asks about the tactile sensation of the cloth in each condition and consists of four questions: roughness, thickness, hardness, and warmth, each answered on a 5-point Likert scale, "yes" or "no" answer if they can accept the cloth as tactile free answer of how they feel when they touch the cloth, selection of the most similar tactile sensation cloth among the 16 prepared fabrics.

Respondents were asked to answer on a 5-point Likert scale for roughness, thickness, stiffness, and warmth. To reduce the influence of individual differences in responses, they were asked to answer based on four fabrics: jeans, voile, gauze, and towels.
Four fabrics were selected for the study. All were 100\% cotton and were familiar to the participants because of their common use. Each criterion was developed as shown in the Table~\ref{tab:cloth_criteria}. Subjects compared each criterion by touching the fabric and the conductive fabric and wrote down the most appropriate score.

For the similar fabrics question, the 16 fabrics are: 1: Marquisette, 2: Cashmere, 3: Satin, 4: Milanese, 5: Faille, 6: Viyella, 7: Metallic Tone Cloth, 8: Georgette, 9: Dungaree, 10: Quilting, 11: Paisley, 12: Velveteen, 13: Chenille, 14: Cambric, 15: Cord Weave, 16: Blister Cloth. Subjects touched each fabric and the conductive fabric and selected the one they felt was most similar.

\begin{table*}
  \caption{Results of the statistical analysis.}
  \label{tab:commands}
  \begin{tabular}{r r r r r r r r }
    \toprule
    \multicolumn{2}{ c }{roughness} & \multicolumn{2}{ c }{stiffness}   & \multicolumn{2}{ c }{thickness} & \multicolumn{2}{ c }{warmth} \\
    \midrule
    Term                        &P value                &Term                           &P value      & Term                        &P value                &Term                           &P value    \\
    \midrule
    Voltage                     &$7.76 \times 10^{-5}$  &   Voltage                     &$0.317$      & Voltage                     &$0.044$                &   Voltage                     &$0.086$    \\
    Frequency                   &$0.263$                &   Frequency                   &$0.281$      & Frequency                   &$0.958$                &   Frequency                   &$0.085$    \\
    Voltage $\times$ Frequency  &$0.777$                &   Voltage $\times$ Frequency  &$0.813$      & Voltage $\times$ Frequency  &$0.901$                &   Voltage $\times$ Frequency  &$0.801$    \\
    \bottomrule
  \end{tabular}
\end{table*}

\subsection{Result}
Through these experiments, we have confirmed that this device changes the tactile sensation by controlling the electrostatic adsorption force, and that even if the tactile sensation changes, the softness and flexibility as a cloth are maintained to some extent. For the tactile sensation, we investigated in detail whether the tactile sensation changed, the effects of voltage and frequency on the tactile sensation, and the characteristics of the presented tactile sensation. For the cloth properties, we investigated whether it was acceptable as a cloth and what kind of cloth it resembled.

\subsubsection{Displayed Tactile Sensation}

% \subparagraph{\textbf{Change in Tactile Sensation}}
% \textbf{Change in Tactile Sensation}
% \paragraph{\textbf{Change in Tactile Sensation}}
\paragraph{Change in Tactile Sensation}
From the experiment, we found that the tactile sensations generally changed when the voltage and frequency were changed; of the nine tactile sensations displayed, the subjects responded that they felt an average of seven different tactile sensations. The reason why all the participants did not perceive any of the nine tactile sensations may be because they judged the low voltage of 100 V to be the same as that of the original cloth.

% \subparagraph{\textbf{Effects of Voltage and Frequency on Tactile Sensation}}
% \textbf{Effects of Voltage and Frequency on Tactile Sensation}
\paragraph{Effects of Voltage and Frequency on Tactile Sensation}
Table~\ref{tab:commands} shows the results. We found that voltage has a statistically significant effect on roughness.  In other conditions, voltage and frequency and their interaction are not statistically significant. However, the p-value for voltage on thickness and the p-value for frequency on heat are somewhat low, suggesting that voltage may have some effect on thickness and frequency may have some effect on heat.

\begin{figure*}[tb]
    \centering
    \includegraphics[width=\linewidth]{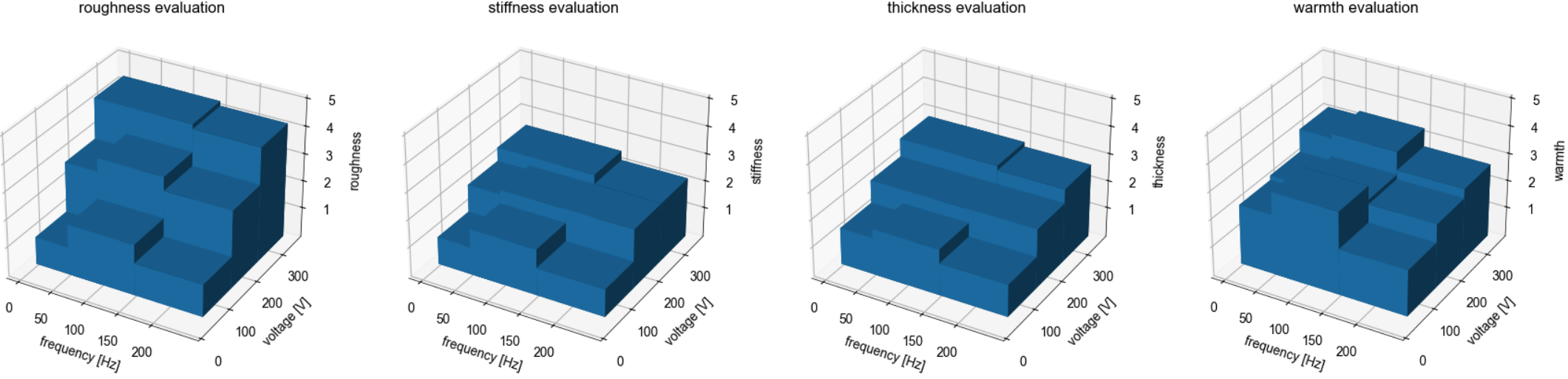}
    \caption{Distribution of means of Likert scale responses for each voltage and frequency.}~\label{fig:bar_plot}
\end{figure*}

% \subparagraph{\textbf{Detailed Property of the Tactile Sensation}}
% \textbf{Detailed Property of the Tactile Sensation}
\paragraph{Detailed Property of the Tactile Sensation}
Figure~\ref{fig:bar_plot} shows the average of the Likert scale responses for roughness, stiffness, thickness, and warmth for each of the nine states where voltage was applied to each characteristic. It can be seen that voltage tends to affect roughness, stiffness, and thickness. On the other hand, frequency tends to have little effect on roughness, stiffness, thickness, and warmth.

Free response also confirms that the tactile sensations presented by the system tend to be of varying roughness, hardness, and thickness with voltage: at 100 V, "light and soft," "kind of slippery, smooth and very soft," "feels like silk," "smooth," "soft," smooth and cool," not much different from the first cloth. Back to the beginning; at 200V it felt tight and rubbery. Feels thicker. Rough". A little rough. The responses for 300 V were rough, "Very rough," "The roughness has increased considerably," "Feels like ribbon fabric," and "Thick and hard.

\subsubsection{Properties as Cloth}
% \textbf{Is it acceptable as a fabric?}
% \subparagraph{\textbf{Is It Acceptable as a Cloth?}}
% \subsubsubsection{Is it acceptable as a fabric?}

% \textbf{Is It Acceptable as a Cloth?}
\paragraph{Is It Acceptable as a Cloth?}
All six subjects indicated that the tactile sensations were acceptable as cloth tactile sensations, except for the 300V, 50Hz, 300V, 200Hz condition. This indicates that the system is capable of presenting multiple tactile sensations while generally maintaining the tactile sensation of cloth. On the other hand, two subjects responded that the tactile sensation of the cloth was unacceptable at 300 V, 50 Hz and 300 V, 200 Hz. In their free responses, at 300 V, 50 Hz, "It was the most vibrating I've ever felt when I stroked it" and "'crackling'"; at 300 V, 200 Hz, "The electrical vibration was so strong that it didn't feel like a fabric. It felt like polyester. The responses were as follows. It is believed that the characteristics of electrostatic adsorption were felt more strongly than the characteristics of the fabric, and were not accepted as the tactile sensation of the fabric.

\begin{figure}[t]
  \centering
  \includegraphics[width=\linewidth]{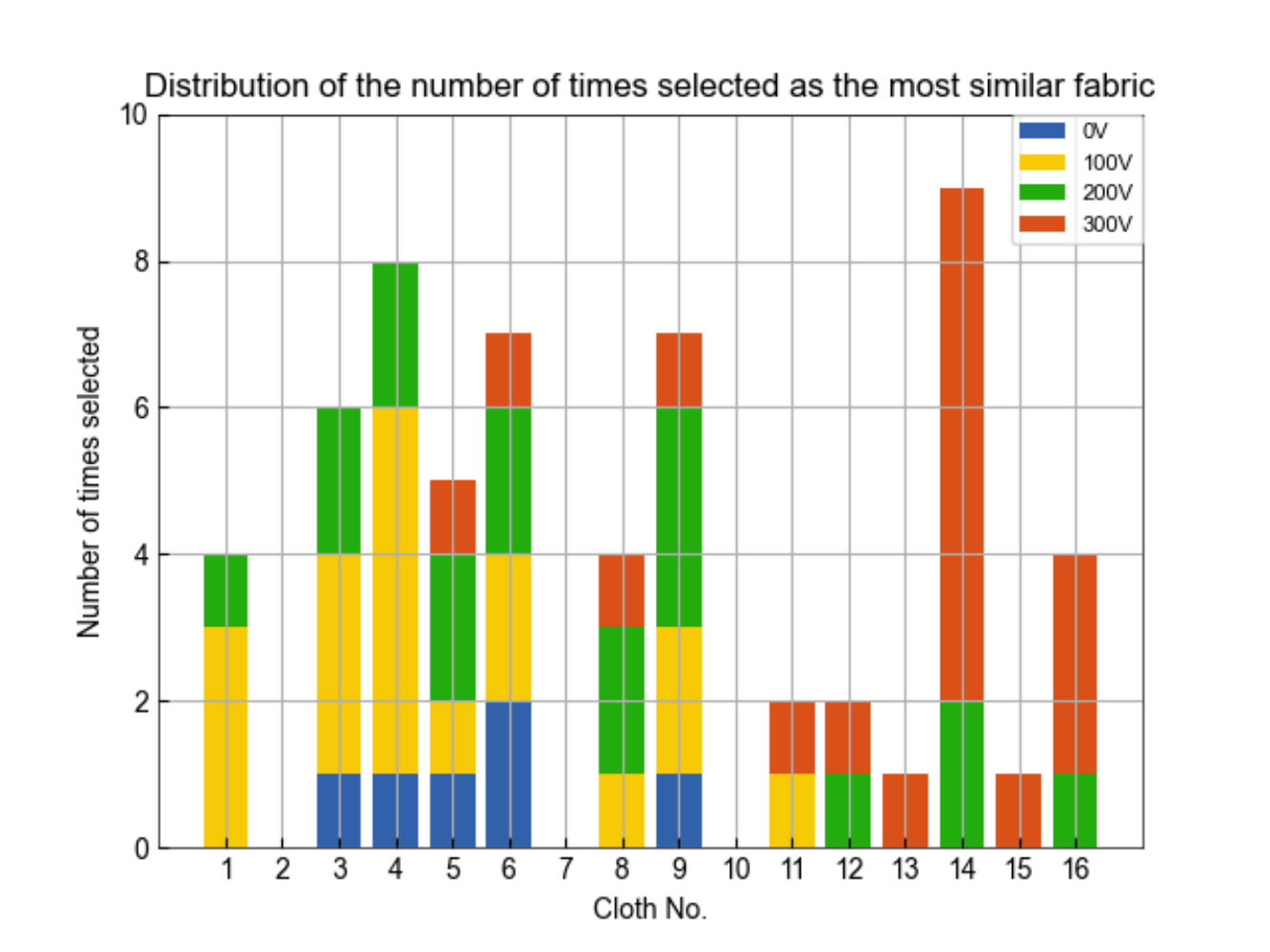}
  \caption{Distribution of the number of times selected as the most similar fabric.}
  \label{fig:dist_similar_cloth}
\end{figure}

% \subparagraph{\textbf{Tactile Similarity to Familiar Cloth}}
% \textbf{Tactile Similarity to Familiar Cloth}
\paragraph{Tactile Similarity to Familiar Cloth}
Figure ~\ref{fig:dist_similar_cloth} shows the distribution of the number of times the cloth was selected as the most similar. 3, 4, 5, 6, and 9 were selected most often, and these are all fabrics that were also selected when the original fabric was not electrified. This indicates that many respondents indicated that the texture was similar to the original cloth. It can also be seen that as the voltage is increased, more fabrics with a strong coarseness, such as 14, are selected. This indicates that the presented tactile impression is strongly influenced by the original cloth, but as the voltage is increased, the roughness changes to become stronger, and it is even possible to present a tactile impression that is different from that of the original cloth. On the other hand, fabrics such as 2, 7, and 10, which were not selected, had a characteristically high weave height. The conductive fabrics used in this study were smooth, and those with many folds and a strong roughness may have been difficult to express.

\section{discussion}

Pinching Tactile Display can dynamically change the tactile sensation of a cloth, and the user can experience multiple tactile sensations with a single device. User tests show that the roughness changes mainly by voltage, the soft and flexible nature of the cloth is maintained, and the tactile sensation of the original conductive cloth is retained, but by increasing the voltage, a tactile sensation different from the original can be presented.
While this system has the potential to remotely verify the tactile qualities of fabrics when buying clothes online, or to be used for interactions in virtual reality spaces to enhance the immersive experience, there is some room for improvement in our system.

The need to wear gloves is cumbersome to experience the system. The gloves used in this study are thin, inexpensive, widely available, and easy to obtain, but not everyone has easy access to such gloves. It is necessary to lower the voltage or substitute another insulator so that the user can touch and feel the system without wearing gloves.

It is possible to increase the roughness of tactile sensations with our system, but reducing it is a challenge.
This suggests that starting with a conductive fabric that inherently has a higher level of roughness may be more advantageous for achieving a wider range of tactile sensations. By selecting such a fabric as the base material, we could potentially improve the system's ability to simulate a wider range of textures.
In terms of waveform manipulation, our study focused primarily on square waves. However, the potential impact of other waveforms, such as sawtooth waves, on tactile sensations remains an area ripe for investigation.Different waveforms may produce different tactile effects, providing a new method for fine-tuning the tactile experience. Exploring these waveforms could lead to more nuanced control over the tactile sensations generated by the system.

There are several things we were not able to explore in this experiment. For example, the gloves may have prevented the subjects from feeling more delicate tactile sensations than when touching with bare hands. Touching the object with the visuals blocked might have allowed for a more purely tactile investigation. The number of subjects was six, and the robustness of the results may be questionable. These are the next points to be verified.

To increase the expressiveness of this device, the creation of specific conductive patterns within the fabric should be explored.
This approach would involve designing and integrating different conductive patterns into the fabric, which could then interact with electrostatic forces to produce different tactile effects.
\section{conclusion}
% In this paper, we presented Pinching Tactile Display, a cloth that changes tactile sensations by electrostatic adsorption. Our evaluation showed that this device can present different tactile sensations of different cloths in one device. In particular, we observed statistically significant differences in roughness by changing the voltage. We found that the tactile sensations that can be represented by this device are similar to the tactile properties of materials such as those used in underwear. In the future, the goal is to allow online shoppers to experience the tactile properties of fabrics before purchasing garments.

In this paper, we proposed Pinching Tactile Display, a cloth that changes tactile sensations by electrostatic adsorption. 
This device can display different tactile sensations of different cloths in one device while maintaining its soft and flexible property. Users can experience multiple tactile sensations with a single device. In addition, this system provides tactile sensation throughout the conductive cloth, so users can explore the tactile sensation by pinching and rubbing the cloth as they normally do when touching cloth. The user test shows that the roughness is mainly changed by the voltage, the soft and flexible nature of the cloth is maintained, and the tactile sensation of the original conductive cloth is maintained, but by increasing the voltage, a tactile sensation different from the original can be displayed.
In the future, we aim to use this technology to enhance human interaction with fabrics in applications ranging from online shopping and wearable technology to immersive virtual reality experiences.

%%
%% The acknowledgments section is defined using the "acks" environment
%% (and NOT an unnumbered section). This ensures the proper
%% identification of the section in the article metadata, and the
%% consistent spelling of the heading.
\begin{acks}
This work was supported by JST Moonshot R\&D Grant JPMJMS2012, JPNP23025 commissioned by the New Energy and Industrial Technology Development Organization (NEDO), and ZOZO NEXT, Inc.
\end{acks}

% This work was supported by JST Moonshot R\&D Grant JPMJMS2012, JST CREST Grant JPMJCR17A3, and the commissioned research by NICT Japan Grant JPJ012368C02901, and ZOZO NEXT, Inc.

%%
%% The next two lines define the bibliography style to be used, and
%% the bibliography file.
\bibliographystyle{ACM-Reference-Format}
% \bibliography{sample-base}
% \bibliography{bib-base}
\bibliography{Source/bib-base}

%%
%% If your work has an appendix, this is the place to put it.
\appendix

\end{document}